\documentclass[prd,aps,tightenlines,a4paper,12pt]{revtex4}

\usepackage{graphicx}
\usepackage{bm}

\usepackage{url}

\newcommand{\OO}[1]{{\mathcal O}(c^{-#1})}

\newcommand{\muas}[0]{\hbox{\rm $\mu$as}}

\newcommand{\ve}[1]{\mbox{\boldmath$#1$}}

\begin{document}

\title{Parametrized post-post-Newtonian analytical solution for light propagation}

\author{Sergei A. \surname{Klioner}, Sven \surname{Zschocke}}

\affiliation{Lohrmann Observatory, Dresden Technical University,
Mommsenstr. 13, 01062 Dresden, Germany}

\begin{abstract}
\begin{center}
\bigskip
\bf
GAIA-CA-TN-LO-SK-002-2

\medskip

issue 2, 24 Februar 2009 
\end{center}

\medskip

An analytical solution for light propagation in the post-post-Newtonian
approximation is given for the Schwarzschild metric in harmonic gauge
augmented by PPN and post-linear parameters $\beta$, $\gamma$ and
$\epsilon$. The solutions of both Cauchy and boundary problem are
given. The Cauchy problem is posed using the initial position of the
photon $\ve{x}_0 = \ve{x}(t_0)$ and its propagation direction
\ve{\sigma} at minus infinity: $\ve{\sigma} = {1\over
c}\,\lim\limits_{t \rightarrow -\infty}\dot{\ve{x}}(t)$. An analytical
expression for the total light deflection is given. The solutions for
$t - t_0$ and $\ve{\sigma}$ are given in terms of boundary conditions
$\ve{x}_0 = \ve{x} (t_0)$ and $\ve{x} = \ve{x}(t)$.
\end{abstract}


\maketitle

\tableofcontents

\newpage

\section{Introduction}

The goal of this note is to derive a rigorous analytical solution for
light propagation in the gravitational field of one spherically
symmetric body in the framework of parametrized post-Newtonian (PPN)
formalism extended by a
non-linear parameter for the component of order $c^{-4}$ in $g_{ij}$.
Our interest to the analytical post-post-Newtonian solution for light
propagation is caused by the inability of the post-Newtonian solution
to predict the light deflection with an accuracy of 1 \muas\ for solar
system objects observed close to Jupiter and Saturn
\citep{KlionerBlankenburg2003}. The maximal errors of the standard
post-Newtonian solution obtained by a comparison of that solution with
numerical integrations of the geodetic equations for light propagation
may in some cases attain 16 \muas\ (the details of these deviations
will be discussed in a subsequent report). Because of this drawback of
the post-Newtonian solution the current implementation
\citep{KlionerBlankenburg2003,Klioner2003b} of the Gaia baseline model
GREM \citep{Klioner2003} uses a kind of time-consuming numerical
inversion procedure to compute the light deflection effect for solar
system bodies if the accuracy below $\sim20$ \muas\ is requested.
Although it does not seem to be a problem for the practical use of the
model, a fully analytical solution for light propagation is certainly
desirable.

The geodetic equation for the light ray in Schwarzschild metric can in
principle be integrated exactly \citep{Chandrasekhar1983}. However,
such an analytical solution is given in terms of elliptic integrals and
is not very suitable for massive calculations. Besides that, only the
trajectory of the photon is readily available from the literature, but
not the position and velocity of a photon as functions of time.
Fortunately, in many cases of interest approximate solutions are
sufficient. The standard way to solve the geodetic equation is the
well-known post-Newtonian approximation scheme. Normally, in current
practical applications of light propagation in the relativistic models
the first post-Newtonian solution is used. Post-post-Newtonian effects
have been also sometimes considered \citep{Hellings1986,Moyer2000}, but
in a way which cannot be called self-consistent since no rigorous
solution in the post-post-Newtonian approximation has been used. Such a
rigorous post-post-Newtonian analytical solution for light propagation
in the Schwarzschild metric of general relativity has been derived
by \citet{Brumberg1987,Brumberg1991}. Although the author has considered the
solution in a class of gauges introducing gauge parameters, this
parametrization does not cover alternative theories of gravity and
therefore, a post-post-Newtonian solution for light propagation within
the PPN formalism and its extension to the second post-Newtonian
approximation is not known. However, it is clearly advantageous to have
such a solution within the PPN formalism. The goal of this note is to
repeat the post-post-Newtonian solution of \citet{Brumberg1987} and to
extend it for the boundary problem.

\section{Differential equations of light propagations and their integral}

The purpose of this Section is to derive the differential equations of light
propagations with PPN and post-linear parameters. We also find an integral of these equations
from the condition that the solution should represent an isotropic geodetic.

\subsection{Metric tensor}

It is well known that in harmonic gauge

\begin{equation}
\label{harmonic-conditions}
\frac{\partial \left( \sqrt{- g} \, g^{\alpha \beta} \right)}{\partial x^{\beta}}= 0
\label{harmonic_5}
\end{equation}

\noindent
the components of the covariant metric tensor of the Schwarzschild
solution are given by

\begin{eqnarray}
g_{00} &=& - \, \frac{1-a}{1+a} ,
\nonumber\\
g_{0i} &=& 0 ,
\nonumber\\
g_{ij} &=& \left(1 + a \right)^2 \, \delta_{i j} \, + \,
\frac{a^2}{x^2} \, \frac{1+a}{1-a} \, x^i \, x^j
\label{exact_5}
\end{eqnarray}

\noindent
with

\begin{equation}
a={m\over x},
\end{equation}

\noindent
where $m={GM\over c^2}$ is the Schwarzschild radius of a body with mass
$M$, $G$ is the Newtonian gravitational constant, $c$ is the light
velocity, and $x= |\ve{x}| = \sqrt{\delta_{i j} \, x^i \, x^j} =
\sqrt{\left(x^1\right)^2 + \left(x^2\right)^2 + \left(x^3\right)^2}$ is
the Euclidean absolute value of vector $\ve{x}$. Expanding this metric
in powers of $c^{-1}$, retaining only the terms relevant for the
post-post-Newtonian solution for the light propagation, and introducing
the PPN parameters $\beta$ and $\gamma$ \citep{Will1993} and the
post-linear parameter $\epsilon$ one gets

\begin{eqnarray}\label{g00}
g_{00}&=&-1+2\,a-2\,\beta\,a^2+\OO6,
\nonumber
\\
\label{g0i}
g_{0i}&=&0,
\nonumber
\\
\label{gij}
g_{ij}&=&\delta_{ij}+2\,\gamma\,a\,\delta_{ij}
+\epsilon\,\left(\delta_{ij}+{x^i\,x^j\over x^2}\right)\,a^2+\OO6 .
\label{ppN_covariant}
\end{eqnarray}

\noindent
In general relativity one has $\beta=\gamma=\epsilon=1$.
Parameter $\epsilon$ should be considered as a formal way to trace, in
the following calculations, the terms coming from the terms $c^{-4}$ in
$g_{ij}$. No physical meaning of $\epsilon$ is claimed in this report.
However, this parameter is equivalent to parameter $\Lambda$ of
\citet{richtermatzner1,richtermatzner2,richtermatzner3} and parameter $\epsilon$ of
\citet{EpsteinShapiro1980}.

The corresponding contravariant components of metric tensor can be
deduced from (\ref{ppN_covariant}) and are given by

\begin{eqnarray}
g^{00}&=&-1-2\,a+2\,(\beta-2)\,a^2+\OO6,
\nonumber
\\
g^{0i}&=&0,
\nonumber
\\
g^{ij}&=&\delta_{ij}-2\,\gamma\,a\,\delta_{i j}
+\left((4\gamma^2-\epsilon)\,\delta_{ij}
-\epsilon\,{x^i\,x^j\over x^2}\right)\,a^2 + \OO6.
\label{ppN_contravariant}
\end{eqnarray}

\noindent
The determinant of metric tensor reads

\begin{eqnarray}
\label{detg}
g&=&-1-2\,(3\,\gamma-1)\,a
-2\,(\beta+2\,\epsilon+6\,\gamma\,(\gamma-1))\,a^2+\OO6,
\end{eqnarray}

\begin{eqnarray}
\label{sqrt-detg}
\sqrt{-g}&=&1+(3\,\gamma-1)\,a
+(2\,\beta+4\,\epsilon-1+3\,\gamma\,(\gamma - 2))\,a^2 + \OO6.
\end{eqnarray}

\noindent
Metric (\ref{gij}) is obviously harmonic for $\gamma=\beta=\epsilon=1$
since the harmonic conditions (\ref{harmonic-conditions}) take the form

\begin{eqnarray}
{\left(\sqrt{-g }\,g^{0\alpha}\right)}_{,\,\alpha}&=&0,
\nonumber
\\
{\left(\sqrt{-g}\,g^{i\alpha}\right)}_{,\,\alpha}&=&
(1-\gamma)\,{a\,x^i\over x^2}
+((1+\gamma)^2-2\beta-2\epsilon)\,{a^2 \, x^i \over x^2}
+\OO6.
\end{eqnarray}

\subsection{Christoffel symbols}

The Christoffel symbols of second kind, defined as

\begin{equation}
\Gamma^{\mu}_{\alpha \beta} = \frac{1}{2}\,g^{\mu\nu}
\left( \frac{\partial g_{\nu \alpha} }{\partial x^{\beta}}
+ \frac{\partial g_{\nu \beta} }{\partial x^{\alpha}} -
\frac{\partial g_{\alpha \beta} }{\partial x^{\nu}} \right) ,
\label{christoffel}
\end{equation}

\noindent
can be derived from metric (\ref{ppN_covariant})--(\ref{ppN_contravariant}):

\begin{eqnarray}
\label{G000}
\Gamma^0_{\ 00} &=& 0 ,
\\
\label{G00i}
\Gamma^0_{\ 0i} &=& {a\,x^i\over x^2} + (1-\beta)\,{2\,a^2\,x^i\over x^2}+ \OO6 ,
\\
\label{G0ik}
\Gamma^0_{\ ik} &=& 0 ,
\\
\label{Gi00}
\Gamma^i_{\ 00}&=&{a\,x^i\over x^2}
- (\beta + \gamma) \, {2\, a^2\, x^i \over x^2}+\OO6 ,
\\
\label{Gi0k}
\Gamma^i_{\ 0k} &=& 0 ,
\\
\label{Gikl}
\Gamma^i_{\ kl} &=& \gamma\,\left(x^i\,\delta_{kl}-x^k\,\delta_{il}-x^l\,
\delta_{ik}\right)\,{a \over x^2}
\nonumber\\
&& +
\left(2\,(\epsilon-\gamma^2)\,x^i\,\delta_{kl}-(\epsilon-2\,\gamma^2)\,
\left(x^k\,\delta_{il}+x^l\,\delta_{ik}\right)
-2\,\epsilon\,{x^i\,x^k\,x^l\over x^2}
\right)\, {a^2 \over x^2}
+\OO6 .
\end{eqnarray}

\subsection{Isotropic condition for the null geodetic}

From now on $x^{\alpha}$ denote the coordinates of a photon, $x^i$
denote the spatial coordinates of the photon, and $x=|\ve{x}|$ is the
distance of the photon from the gravitating body that is situated at
the origin of the used reference system. The conditions that a photon
follows an isotropic geodetic can be formulated as an equation for the
four components of the coordinate velocity of that photon $\dot
x^{\alpha}$:

\begin{equation}
g_{\alpha\beta} \, \frac{d \, x^{\alpha}}{d \, \lambda}\,
\frac{d \, x^{\beta}}{d \, \lambda} = 0,
\label{isotropic_5}
\end{equation}

\noindent
$\lambda$ being the canonical parameter, or

\begin{equation}
\label{isotropic1}
g_{00}+{2\over c}\,g_{0i}\,\dot x^i+{1\over c^2}\,g_{ij}\,\dot x^i\,
\dot x^j=0 \, ,
\label{isotropic_10}
\end{equation}

\noindent
where a dot denotes the derivative with respect to coordinate time $t$.
Eq. (\ref{isotropic_10}) is a first integral of motion for the differential
equation for light propagation and must be valid for any point of an
isotropic geodetic. Substituting the ansatz $\dot{\ve{x}}=c\,s\,\ve{\mu}$,
where $\ve{\mu}$ is a unit coordinate direction of light propagation
($\ve{\mu}\cdot\ve{\mu}=1$) and $s=|\dot{\ve{x}}|/c$, into Eq.~(\ref{isotropic_10})
one gets for metric (\ref{ppN_covariant})

\begin{equation}
\label{s-ppN}
s=1-(1+\gamma) \, a + {1\over 2}\,\left(-1+2\,\beta-\epsilon+\gamma\,
(2+3\gamma)-\epsilon\,{\left({\ve{\mu}\cdot\ve{x}\over x}\right)}^2\right) \, a^2
+ {\cal O} (c^{-6}) \, .
\label{isotropic_25}
\end{equation}

\noindent
This formula allows one to compute the absolute value of coordinate
velocity of light in the chosen reference system if the position of the
photon $x^i$ and the coordinate direction of its propagation $\mu^i$
are given.

\subsection{Equations of light propagation}

The geodetic equation is given by

\begin{equation}
\frac{d^2 x^{\mu}}{d \lambda^2} + \Gamma^{\mu}_{\alpha \beta}\,
\frac{d x^{\alpha}}{d \lambda}\,\frac{d x^{\beta}}{d \lambda} = 0 \, .
\label{geodetic_5}
\end{equation}

\noindent
Using the relation of the affine parameter $\lambda$ and the coordinate time
$t = x^0 / c$

\begin{equation}
\frac{d^2 t}{d \lambda^2}  + {1\over c}\, \Gamma^{0}_{\alpha \beta}
\frac{d x^{\alpha}}{d \lambda}\,\frac{d x^{\beta}}{d \lambda} = 0
\label{geodetic_10}
\end{equation}

\noindent
one can reparametrize the geodetic equation by coordinate time $t$,
\begin{eqnarray}
\ddot x^i &=&
-c^2\,\Gamma^i_{\ 00}
-2\,c\,\Gamma^i_{\ 0j}\,\dot x^j
-\Gamma^i_{jk}\,\dot x^j\,\dot x^k
+\dot x^i\,
\left(
c\,\Gamma^0_{\ 00}
+2\,\Gamma^0_{\ 0j}\,\dot x^j
+{1\over c}\,\Gamma^0_{jk}\,\dot x^j\,\dot x^k
\right)\,.
\label{exact_20}
\end{eqnarray}

Inserting the Christoffel symbols (\ref{G000})--(\ref{Gikl}), one gets the following
equations of light propagation in post-post-Newtonian approximation,
\begin{eqnarray}
\label{geodetic-ppN}
\ddot{\ve{x}} &=&
- \left(c^2+\gamma\,\dot{\ve{x}}\cdot\dot{\ve{x}}\right)\,{a\,\ve{x}\over x^2}
+2\,(1+\gamma)\,{a\,\dot{\ve{x}}\,(\dot{\ve{x}}\cdot\ve{x})\over x^2}
\nonumber
\\
&&
+2\,\left((\beta+\gamma)\,c^2+(\gamma^2-\epsilon)\,
(\dot{\ve{x}}\cdot\dot{\ve{x}})\right)\,{a^2\,\ve{x}\over x^2}
+2\,\epsilon\,{a^2\,\ve{x}\,(\dot{\ve{x}}\cdot\ve{x})^2\over x^4}
\nonumber
\\
&&
+2\,(2(1-\beta)+\epsilon-2\,\gamma^2)\,
{a^2\,\dot{\ve{x}}\,(\dot{\ve{x}}\cdot\ve{x})\over x^2} + \OO4 \,.
\end{eqnarray}

\noindent
Here, for estimating the analytical order of smallness of the terms
we take into account that $|\dot{\ve{x}}| = {\cal O}(c)$. Using
(\ref{s-ppN}) and $\dot{\ve{x}}\cdot\dot{\ve{x}}=c^2\,s^2$
one can simplify (\ref{geodetic-ppN}) to get

\begin{eqnarray}
\label{geodetic-ppN-final}
\ddot{\ve{x}} &=&
-(1+\gamma)\,c^2\,{a\,\ve{x}\over x^2}
+2\,(1+\gamma)\,{a\,\dot{\ve{x}}\,(\dot{\ve{x}}\cdot\ve{x})\over x^2}
\nonumber
\\
&&
+2\,c^2\,\left(\beta-\epsilon+2\,\gamma\,(1+\gamma)\right)\,
{a^2\,\ve{x}\over x^2} + 2\,\epsilon\,{a^2\,\ve{x}\,(\dot{\ve{x}}\cdot\ve{x})^2\over x^4}
\nonumber
\\
&&
+2\,(2(1-\beta)+\epsilon-2\,\gamma^2)\,
{a^2\,\dot{\ve{x}}\,(\dot{\ve{x}}\cdot\ve{x})\over x^2} + \OO4 \,.
\end{eqnarray}

\subsection{Equations of light propagation with additional empirical
parameter $\alpha$}

For our purposes it is advantageous to have one more additional
parameter that can be used to trace terms in the folowing calculations
which come from the post-post-Newtonian terms in the equations of
motion of a photon. We denote this parameter by $\alpha$ and introduce
it in the above equation simply as a factor for all the
post-post-Newtonian terms:
\begin{eqnarray}
\ddot{\ve{x}} &=&
-(1+\gamma)\,c^2\,{a\,\ve{x}\over x^2}
+2\,(1+\gamma)\,{a\,\dot{\ve{x}}\,(\dot{\ve{x}}\cdot\ve{x})\over x^2}
\nonumber
\\
&&
+2\,c^2\,\alpha\,\left(\beta-\epsilon+2\,\gamma\,
(1+\gamma)\right)\,{a^2\,\ve{x}\over x^2}
+2\,\alpha\,\epsilon\,{a^2\,\ve{x}\,(\dot{\ve{x}}\cdot\ve{x})^2\over x^4}
\nonumber
\\
&&
+2\,\alpha\,(2(1-\beta)+\epsilon-2\,\gamma^2)\,
{a^2\,\dot{\ve{x}}\,(\dot{\ve{x}}\cdot\ve{x})\over x^2}
+\OO4\,.
\label{geodetic-ppN-final-with-alpha}
\end{eqnarray}

\noindent
Setting $\alpha=0$ in the solution of Eq.
(\ref{geodetic-ppN-final-with-alpha}) one can get formally a
second-order solution for the post-Newtonian equations of light
propagation.

\section{Initial value problem}

Let us now solve analytically an initial value problem
for the derived equations.

\subsection{Analytical post-post-Newtonian solution}

For two initial conditions
\begin{eqnarray}
\ve{x}_0 &=& \ve{x}(t_0)\,,
\nonumber\\
\ve{\sigma} &=& \lim\limits_{t \rightarrow - \infty}\,\frac{\dot{\ve{x}}(t)}{c}\,,
\label{cauchy_5}
\end{eqnarray}

\noindent
using the same approach as was used by \citet{Brumberg1987}, one gets a
formal solution $\ve{x}(t)$ and $\dot{\ve{x}}(t)$ of Eq.
(\ref{geodetic-ppN-final-with-alpha}):
\begin{eqnarray}
\label{rN-rpN-rppN}
{1\over c}\,\dot{\ve{x}}_N&=&\ve{\sigma},
\label{brumberg_5}
\\
\ve{x}_N&=&\ve{x}_0+c\,(t-t_0)\,\ve{\sigma},
\label{brumberg_10}
\\
{1\over c}\,\dot{\ve{x}}_{\rm pN}&=&\ve{\sigma}+m\,\ve{A}_1(\ve{x}_N),
\label{brumberg_15}
\\
\ve{x}_{\rm pN}&=&\ve{x}_N+m\,\left(\ve{B}_1(\ve{x}_N)-\ve{B}_1(\ve{x}_0)\right),
\label{brumberg_20}
\\
{1\over c}\,\dot{\ve{x}}_{\rm ppN}&=&\ve{\sigma}+m\,\ve{A}_1(\ve{x}_{\rm pN})
+m^2\,\ve{A}_2(\ve{x}_N),
\label{brumberg_25}
\\
\ve{x}_{\rm ppN}&=&\ve{x}_N+m\,\left(\ve{B}_1(\ve{x}_{\rm pN})-\ve{B}_1(\ve{x}_0)\right)
+m^2\,\left(\ve{B}_2(\ve{x}_N)-\ve{B}_2(\ve{x}_0)\right),
\label{brumberg_30}
\end{eqnarray}

\noindent
with
\begin{eqnarray}
\ve{A}_1(\ve{x}) &=& -(1+\gamma)\,
\left({{\ve{\sigma}\times(\ve{x}\times\ve{\sigma})\over x
(x -\ve{\sigma}\cdot\ve{x})}+{\ve{\sigma}\over x}}\right),
\label{brumberg_35}
\\
\ve{B}_1(\ve{x})&=&-(1+\gamma)\,
\left({\ve{\sigma}\times(\ve{x}\times\ve{\sigma})\over x
-\ve{\sigma}\cdot\ve{x}}
+\ve{\sigma}\,\log {( x +\ve{\sigma}\cdot\ve{x})}\right),
\label{brumberg_40}
\\
\label{A_2a}
\ve{A}_2(\ve{x})&=&
-{1\over 2}\,\alpha\,\epsilon\,{\ve{\sigma}\cdot\ve{x}\over x^4}\,\ve{x}
+2\,(1+\gamma)^2\,{\ve{\sigma}\times(\ve{x}\times\ve{\sigma})\over x^2\,
\left( x - \ve{\sigma}\cdot\ve{x}\right)}
+(1+\gamma)^2\,{\ve{\sigma}\times(\ve{x}\times\ve{\sigma})\over x \,
{\left(x -\ve{\sigma}\cdot\ve{x}\right)}^2}
\nonumber
\\
&&
-(1+\gamma)^2\,{\ve{\sigma}\over x\,\left(x-\ve{\sigma}\cdot\ve{x}\right)}
+\left(2(1-\alpha+\gamma)\,(1+\gamma)+\alpha\,
\beta-{1\over 2}\,\alpha\,\epsilon\right)\,{\ve{\sigma}\over x^2}
\nonumber
\\
&&
-{1\over 4}\,\left(8\,(1+\gamma-\alpha\,\gamma)\,(1+\gamma)-4\,\alpha\,\beta+3\,\alpha\,\epsilon\right)\,
\left(\ve{\sigma}\cdot\ve{x}\right)\,
{\ve{\sigma}\times(\ve{x}\times\ve{\sigma})\over x^2\,|\ve{\sigma}\times\ve{x}|^2}
\nonumber
\\
&&
-{1\over 4}\,\left(8\,(1+\gamma-\alpha\,\gamma)\,(1+\gamma)-4\,\alpha\,\beta+3\,\alpha\,\epsilon\right)\,
{\ve{\sigma}\times(\ve{x}\times\ve{\sigma})\over |\ve{\sigma}\times\ve{x}|^3}\,
\left( \pi - \delta (\ve{\sigma}, \ve{x}) \right)
\label{brumberg_45}
\\
\ve{B}_2(\ve{x})&=&
-(1+\gamma)^2\,{\ve{\sigma}\over x - \ve{\sigma}\cdot\ve{x}}
+(1+\gamma)^2\,{\ve{\sigma}\times(\ve{x}\times\ve{\sigma})\over {\left( x - \ve{\sigma}\cdot\ve{x}\right)}^2}
+{1\over 4}\,\alpha\,\epsilon\,{\ve{x}\over x^2}
\nonumber
\\
&&
-{1\over 4}\,\alpha\,\left(8\,(1+\gamma)-4\,\beta+3\,\epsilon\right)\,{\ve{\sigma}\over |\ve{\sigma}\times\ve{x}|}\,
\left( \frac{\pi}{2} - \delta (\ve{\sigma}, \ve{x}) \right)
\nonumber
\\
&&
-{1\over 4}\,\left(8\,(1+\gamma-\alpha\,\gamma)\,(1+\gamma)-4\,\alpha\,\beta+3\,\alpha\,\epsilon\right)\,\left(\ve{\sigma}\cdot\ve{x}\right)\,
{\ve{\sigma}\times(\ve{x}\times\ve{\sigma})\over |\ve{\sigma}\times\ve{x}|^3}\,
\left( \pi - \delta (\ve{\sigma}, \ve{x}) \right),
\nonumber\\
\label{brumberg_50}
\end{eqnarray}
\noindent
or, alternatively, for $\ve{B}_1$ and $\ve{B}_2$
\begin{eqnarray}
\label{B_1a}
\ve{B}_1(\ve{x})&=&-(1+\gamma)\,
\left({\ve{\sigma}\times(\ve{x}\times\ve{\sigma})\over x -
\ve{\sigma}\cdot\ve{x}}
-\ve{\sigma}\,\log {(x-\ve{\sigma}\cdot\ve{x})}\right),
\\
\label{B_2}
\ve{B}_2(\ve{x})&=&
+(1+\gamma)^2\,{\ve{\sigma}\over x -\ve{\sigma}\cdot\ve{x}}
+(1+\gamma)^2\,{\ve{\sigma}\times(\ve{x}\times\ve{\sigma})\over
{\left(x-\ve{\sigma}\cdot\ve{x}\right)}^2}
+{1\over 4}\,\alpha\,\epsilon\,{\ve{x}\over x^2}
\nonumber
\\
&&
-{1\over 4}\,\alpha\,\left(8\,(1+\gamma)-4\,\beta+3\,\epsilon\right)\,{\ve{\sigma}\over |\ve{\sigma}\times\ve{x}|}\,
\left( \frac{\pi}{2} - \delta (\ve{\sigma}, \ve{x}) \right)
\nonumber
\\
&&
-{1\over 4}\,\left(8\,(1+\gamma-\alpha\,\gamma)\,(1+\gamma)-4\,\alpha\,\beta+3\,\alpha\,\epsilon\right)\,\left(\ve{\sigma}\cdot\ve{x}\right)\,
{\ve{\sigma}\times(\ve{x}\times\ve{\sigma})\over |\ve{\sigma}\times\ve{x}|^3}\,
\left( \pi - \delta (\ve{\sigma}, \ve{x}) \right).
\nonumber\\
\end{eqnarray}

\noindent
Here $\delta (\ve{a}, \ve{b})$ is the
angle between two arbitrary vectors $\ve{a}$ and $\ve{b}$. Clearly, for an angle between two vectors one has $0 \le \delta (\ve{a}, \ve{b}) \le \pi$. The angle $\delta
(\ve{a}, \ve{b})$ can be computed in many ways, for example as $\delta
(\ve{a}, \ve{b}) = \arccos \displaystyle{\ve{a} \cdot \ve{b}\over a\,b}$.
With these definitions the solution of
(\ref{geodetic-ppN-final-with-alpha}) reads
\begin{eqnarray}
\ve{x} &=& \ve{x}_{\rm ppN} + {\cal O} (c^{- 6}) ,
\nonumber\\
\frac{1}{c} \dot{\ve{x}} &=& \frac{1}{c} \dot{\ve{x}}_{\rm ppN} + {\cal O} (c^{- 6}) .
\end{eqnarray}

It is easy to check that the solution for coordinate velocity of light
$\dot{\ve{x}}_{\rm ppN}$ satisfies the integral (\ref{s-ppN}). In order to demonstrate
this fact, it is important to understand that position $\ve{x}$ in (\ref{s-ppN}) lies on
the trajectory of the photon and must be therefore considered as $\ve{x}_{\rm pN}$ in
the post-Newtonian terms and as $\ve{x}_{\rm N}$ in the post-post-Newtonian terms of
(\ref{brumberg_25}).

\subsection{Total light deflection}

In order to derive the total light deflection, we have to consider the
limits of coordinate light velocity $\dot{\ve{x}}$ for $t \rightarrow
\pm \infty$:
\begin{eqnarray}
\lim_{t\to-\infty} {1\over c}\,\dot{\ve{x}}(t) &=& \ve{\sigma},
\\
\lim_{t\to+\infty} {1\over c} \, \dot{\ve{x}} (t) &\equiv& \ve{\nu}
\nonumber\\
&=& \ve{\sigma}  -  2\,(1 + \gamma)\,m\,
\frac{\ve{\sigma} \times (\ve{x}_0 \times \ve{\sigma})}{|\ve{x}_0 \times \ve{\sigma}|^2}
 -  2 \, (1 + \gamma)^2\,  m^2\,  \frac{\ve{\sigma}}
{|\ve{x}_0 \times \ve{\sigma}|^2}
\nonumber\\
&& -  \frac{1}{4}\, \pi\,
\left(8 (1 + \gamma - \alpha \, \gamma) (1 + \gamma)
- \, 4  \alpha \, \beta + 3 \alpha \, \epsilon \right)\,  m^2\,
\frac{\ve{\sigma} \times (\ve{x}_0 \times \ve{\sigma})}{|\ve{x}_0 \times \ve{\sigma}|^3}
\nonumber\\
&& +  2\, (1 + \gamma)^2\,  m^2  \,\left( x_0  +  \ve{\sigma} \cdot
\ve{x}_0 \right)  \frac{\ve{\sigma} \times (\ve{x}_0 \times \ve{\sigma})}
{|\ve{x}_0 \times \ve{\sigma}|^4} +\OO6.
\label{deflection_5}
\end{eqnarray}

Accordingly, the total light deflection reads
\begin{eqnarray}
|\ve{\sigma} \times \ve{\nu}| &=& 2 \, (1 + \gamma) \, m \,
\frac{1}{|\ve{x}_0 \times \ve{\sigma}|} \, - \,
2 \, (1 + \gamma)^2 \, m^2 \, ( x_0 \, + \ve{\sigma} \cdot \ve{x}_0) \,
\frac{1}{|\ve{x}_0 \times \ve{\sigma}|^3}
\nonumber\\
&& + \frac{1}{4}
\left(8 (1 + \gamma - \alpha \, \gamma) (1 + \gamma)
- \, 4 \, \alpha \, \beta + 3 \alpha \, \epsilon \right)\,\pi \, m^2 \,
\frac{1}{|\ve{x}_0 \times \ve{\sigma}|^2} +\OO6\,.
\label{deflection_10}
\end{eqnarray}

\noindent
Eq.~(\ref{deflection_10}) defines the sine of the angle of total light
deflection in post-post-Newtonian approximation. The first term in
(\ref{deflection_10}) is the post-Newtonian expression of total light
deflection $\sim4 m / d$, where $d = |\ve{x}_0 \times \ve{\sigma}|$
being the impact parameter for the light trajectory. The other two
terms are the post-post-Newtonian corrections. Note that although the
total light deflection $|\ve{\sigma} \times \ve{\nu}|$ is a
coordinate-independent quantity, $\ve{x}_0$ and, therefore, the impact
parameter $d = |\ve{x}_0 \times \ve{\sigma}|$ are coordinate-dependent.
In order to compare Eq.~(\ref{deflection_10}) with results expressed through
the coordinate-independent impact parameter $d^\prime$ one has to find a
transformation between that impact parameter and $d$. The coordinate-independent
impact parameter is defined as a limit

\begin{equation}
\label{d-prime-definition}
d^\prime = \lim_{t\to-\infty}|\ve{x}(t)\times\ve{\sigma}|
 = \lim_{t\to+\infty}|\ve{x}(t)\times\ve{\nu}| \, .
\end{equation}

\noindent
Substituting the post-Newtonian coordinates (\ref{brumberg_20}) of the
photon $\ve{x}(t)$ into this definition one gets

\begin{equation}
\label{d-prime-d}
d^\prime=d+(1+\gamma)\,m\,{x_0+\ve{\sigma}\cdot\ve{x}_0\over d}+{\cal O}(c^{-4})\,.
\end{equation}

\noindent
It is now clear that the second term in the right-hand side of
(\ref{deflection_10}) just ``corrects'' the main post-Newtonian term
converting it to $2(1+\gamma)m/d^\prime$. Using $d^\prime$ one can write
(\ref{deflection_10}) as
\begin{eqnarray}
|\ve{\sigma} \times \ve{\nu}| &=& 2 \, (1 + \gamma) \,
\frac{m}{d^\prime}
+ \frac{1}{4}\,
\left(8 (1 + \gamma - \alpha \, \gamma) (1 + \gamma)
- \, 4 \, \alpha \, \beta + 3 \alpha \, \epsilon \right)\,\pi\,
\frac{m^2}{d^{\prime2}} +\OO6\,.
\label{full-deflection-d-prime}
\end{eqnarray}

\noindent
This result with $\alpha=1$ coincides with Eq. (4) of
\citet{EpsteinShapiro1980} and also agrees with the results of
\citet{richtermatzner1}, \citet{Cowling1984} and \citet{Brumberg1987}
in the corresponding limits.

\section{Boundary problem}

For practical modeling of observations of solar system objects it is not sufficient to
consider the initial value problem for light propagation. Two-point boundary value problem
is of interest here. Let us consider that the light ray must propagate between two points
being emitted at a position $\ve{x}_0$ at time moment $t_0$ and received
at position $\ve{x}$ at a time moment $t$
\begin{eqnarray}
\ve{x}_0 &=& \ve{x}(t_0) \,,
\nonumber\\
\ve{x} &=& \ve{x}(t) \,.
\label{boundary_5}
\end{eqnarray}

\noindent
Initial time moment $t_0$ can be considered here to be known (although
for any stationary metric like that considered here this moment plays no
role), but the final moment $t$ is unknown. We also denote
\begin{eqnarray}
\ve{R} &=& \ve{x} - \ve{x}_0 \,, \\
\ve{k} &=& \frac{\ve{R}}{R} \,,
\end{eqnarray}

\noindent
where $R=|\ve{R}|$ is the absolute value. In the following, the
solution of geodetic equation (\ref{geodetic-ppN-final-with-alpha})
will be expressed as a function of the boundary values $\ve{x}_0$ and
$\ve{x}$.

\subsection{Formal expressions}

An iterative solution of (\ref{brumberg_5})--(\ref{brumberg_30}) for
the propagation time $\tau=t-t_0$ and unit direction $\ve{\sigma}$ reads
as follows:
\begin{eqnarray}
c \, \tau  = R &-& m\,\ve{k} \cdot
\left[ \ve{B}_1 (\ve{x}) - \ve{B}_1 (\ve{x}_0)\right]
- m^2\,\ve{k} \cdot \left[ \ve{B}_2 (\ve{x}) - \ve{B}_2 (\ve{x}_0)\right]
\nonumber\\
&+&\frac{m^2}{2\,R}\,   \left|
\ve{k} \times \left( \ve{B}_1 (\ve{x}) - \ve{B}_1 (\ve{x}_0) \right)
\right|^2+{\cal O}(c^{- 6}),
\label{iteration_5}
\end{eqnarray}
\begin{eqnarray}
\ve{\sigma} &=& \ve{k} + m\,\frac{1}{R} \,
\left( \ve{k} \times \left[ \ve{k} \times
(\ve{B}_1 (\ve{x}) - \ve{B}_1 (\ve{x}_0)) \right] \right)
\nonumber\\
&&  + m^2\,\frac{1}{R} \,
\left( \ve{k} \times \left[ \ve{k} \times
(\ve{B}_2 (\ve{x}) - \ve{B}_2 (\ve{x}_0)) \right] \right)
\nonumber\\
&&  + m^2\,\frac{1}{R^2} \,
\left(\ve{B}_1 (\ve{x}) - \ve{B}_1 (\ve{x}_0) \right) \times
 \left[ \ve{k} \times (\ve{B}_1 (\ve{x}) - \ve{B}_1 (\ve{x}_0)) \right]
\nonumber\\
&&  - \frac{3}{2}\, m^2\,\frac{1}{R^2} \,\ve{k}\,
\left| \ve{k} \times \left( \ve{B}_1 (\ve{x}) - \ve{B}_1 (\ve{x}_0) \right)
\right|^2+{\cal O}(c^{- 6}).
\label{iteration_10}
\end{eqnarray}

\noindent
These expressions are still implicit since for the post-post-Newtonian
accuracy one should use the post-Newtonian relation between
$\ve{\sigma}$ and $\ve{k}$ in $\ve{B}_1$ appearing in the post-Newtonian
terms. That relation can be again obtained from (\ref{iteration_10}) by
neglecting all terms of order ${\cal O}(m^2)$. On the contrary, in the terms
in (\ref{iteration_5}) and (\ref{iteration_10}) of the order of ${\cal O}(m^2)$
one can use the Newtonian relation $\ve{\sigma} = \ve{k}$.

\subsection{Propagation time}

Substituting (\ref{brumberg_40}) and (\ref{brumberg_50}) into
(\ref{iteration_5}) one can derive an explicit formula for the time of
light propagation:
\begin{eqnarray}
c \, \tau &=& R \, + \, (1 + \gamma) \, m \, \log
\, \frac{x + x_0 + R}{x + x_0 - R}
\nonumber\\
&& + \, \frac{1}{8} \, \alpha \, \epsilon \, \frac{m^2}{R} \,
\left( \frac{x_0^2 - x^2 - R^2}{x^2} \, + \,
\frac{x^2 - x_0^2 - R^2}{x_0^2} \right)
\nonumber\\
&& + \frac{1}{4} \, \alpha \, \left(8 (1 + \gamma) - 4 \beta + 3 \epsilon\right)\,
m^2 \,
\frac{R}{|{\ve{x}} \times {\ve{x}}_0|} \;\delta (\ve{x} , \ve{x}_0)
\nonumber\\
&& + \, \frac{1}{2}\,(1 + \gamma)^2 \, m^2 \, \frac{R}{|{\ve{x}} \times {\ve{x}}_0|^2}
\, \left( (x - x_0)^2 - R^2 \right) +{\cal O}(c^{- 6}) \,.
\label{iteration_7}
\end{eqnarray}

\noindent
This formula allows one to compute time of light propagation $\tau$ for given boundary
conditions $\ve{x}_0$ and $\ve{x}$. Here we have used that 
$\delta (\ve{k} , \ve{x}_0) - \delta (\ve{k} , \ve{x}) = \delta (\ve{x} , \ve{x}_0)$.

\subsection{Transformation of $\ve{k}$ to $\ve{\sigma}$}

In the same way substituting (\ref{brumberg_40}) and (\ref{brumberg_50}) into
(\ref{iteration_10}) one gets:
\begin{eqnarray}
{\ve{\sigma}} &=& {\ve{k}} \, + \, (1 + \gamma) \, m \, \frac{x - x_0 + R}
{|{\ve{x}} \times {\ve{x}}_0|^2} \, {\ve{k}} \times ({\ve{x}}_0 \times {\ve{x}})
\nonumber\\
&& - \, \frac{(1 + \gamma)^2}{2} \, m^2 \, \frac{(x - x_0 + R)^2}
{|{\ve{x}} \times {\ve{x}}_0|^2} \, {\ve{k}}
\nonumber\\
&& + \,m^2\,
{\ve{k}} \times \left( {\ve{x}}_0 \times {\ve{x}} \right)\,
\Bigg[ - \frac{1}{4} \, \alpha \, \epsilon \,\frac{1}{R^2}\,
\left(\frac{1}{x^2}-\frac{1}{x_0^2}\right)
\nonumber\\
&& + \, \frac{1}{8} \, \left(8 (1 + \gamma - \alpha \, \gamma) (1 + \gamma)
- 4 \, \alpha \, \beta + 3 \alpha \, \epsilon\right)
\frac{1}{|{\ve{x}} \times {\ve{x}}_0|^3}\,
\nonumber\\
&&\phantom{aaaaaaaaaa} \times\,
\Biggl(2R^2 \left(\pi - \delta (\ve{k}, \ve{x}) \right)
+ \left(x^2 - x_0^2- R^2\right) \delta (\ve{x}, \ve{x}_0) \Biggr)
\nonumber\\
&& +\frac{1}{2}\,(1 + \gamma)^2 \, \frac{1}{|{\ve{x}} \times {\ve{x}}_0|^4}
\, (x + x_0) \, (x - x_0 - R) \, (x - x_0 + R)^2  \Bigg]+{\cal O}(c^{- 6}) \,.
\label{k_to_sigma}
\end{eqnarray}

\noindent
This formula allows one to compute $\ve{\sigma}$ for given boundary
conditions $\ve{x}_0$ and $\ve{x}$.

\subsection{Transformation of $\ve{\sigma}$ to $\ve{n}$}

Considering the relativistic model of positional observations (e.g.,
\citet{Klioner2003}) it is clear that a unit direction of light
propagation at the point of reception $\ve{n}$ plays here an important
role. This vector is defined as
\begin{equation}
\ve{n}=\frac{\dot{\ve{x}}(t)}{\left|\dot{\ve{x}}(t)\right|}\,.
\end{equation}

\noindent
Expanding Eq.~(\ref{brumberg_25}) up to order ${\cal O} (c^{- 6})$ leads in
post-post-Newtonian approximation
\begin{equation}
\ve{n}=\ve{\sigma} + m\,\ve{C}_1(\ve{x}_{\rm pN}) + m^2\,\ve{C}_2(\ve{x}_{\rm N})
+ {\cal O}(c^{- 6})
\end{equation}

\noindent
with
\begin{eqnarray}
\ve{C}_1 (\ve{x}) &=& \ve{A}_1 (\ve{x})
\,-\,\ve{\sigma} \left(\ve{\sigma} \cdot \ve{A}_1 (\ve{x})\right)
\, = \, - (1 + \gamma) \frac{\ve{\sigma} \times (\ve{x} \times \ve{\sigma})}
{x \, (x - \ve{\sigma} \cdot \ve{x}) } \,,
\nonumber\\
\ve{C}_2 (\ve{x}) &=& \ve{A}_2 (\ve{x})
\,-\, \ve{A}_1 (\ve{x})\,
\left(\ve{\sigma} \cdot \ve{A}_1 (\ve{x}) \right) \,-\, \frac{1}{2}\,
\ve{\sigma}\,\left(\ve{A}_1(\ve{x})\cdot\ve{A}_1(\ve{x})\right)\,-\,
\ve{\sigma}\,\left(\ve{\sigma} \cdot \ve{A}_2 (\ve{x})\right)
\nonumber\\
&& + \frac{3}{2} \,\ve{\sigma}\,
\left(\ve{\sigma} \cdot \ve{A}_1 (\ve{x}) \right)^2
\nonumber\\
&=& - \frac{1}{2}\,\alpha\,\epsilon\, \frac{\ve{\sigma} \cdot \ve{x}}{x^4}\,
\ve{\sigma} \times (\ve{x} \times \ve{\sigma})
+\,(1 + \gamma)^2\,\frac{\ve{\sigma} \times (\ve{x} \times \ve{\sigma})}
{x^2\,(x - \ve{\sigma} \cdot \ve{x})}
\nonumber\\
&& +(1 + \gamma)^2\,
\frac{\ve{\sigma} \times (\ve{x} \times \ve{\sigma})}
{x\,(x - \ve{\sigma} \cdot \ve{x})^2} \,-\,\frac{1}{2}\,
(1 + \gamma)^2\,\frac{\ve{\sigma}}
{x^2}\,\frac{x + \ve{\sigma} \cdot \ve{x}}{x - \ve{\sigma} \cdot \ve{x}}
\nonumber\\
&& -{1\over 4}\,\left(8\,(1+\gamma-\alpha\,\gamma)\,(1+\gamma)-4\,\alpha\,\beta+3\,\alpha\,\epsilon\right)\,
\left(\ve{\sigma}\cdot\ve{x}\right)\,
{\ve{\sigma}\times(\ve{x}\times\ve{\sigma})\over x^2\,|\ve{\sigma}\times\ve{x}|^2}
\nonumber
\\
&&
-{1\over 4}\,\left(8\,(1+\gamma-\alpha\,\gamma)\,(1+\gamma)-4\,\alpha\,\beta+3\,\alpha\,\epsilon\right)\,
{\ve{\sigma}\times(\ve{x}\times\ve{\sigma})\over |\ve{\sigma}\times\ve{x}|^3}\,
\left( \pi - \delta (\ve{\sigma} , \ve{x}) \right)\,.
\label{n_5}
\end{eqnarray}

\noindent
Noting that $\ve{x}_{\rm pN}=\ve{x}+{\cal O}(c^{- 4})$,
$\ve{x}=\ve{x}_0+R\,\ve{k}$ and using Eq. (\ref{k_to_sigma}) for
$\ve{\sigma}$ in $\ve{C}_1(\ve{x}_{\rm pN})$ we get
\begin{eqnarray}
\ve{n} &=& \ve{\sigma}\,-\,(1 + \gamma)\,m\,
\ve{k} \times ( \ve{x}_0 \times \ve{x} )
\frac{R}{|\, \ve{x} \times \ve{x}_0 \,|^2}\,
\left(1\,+\,\frac{\ve{k} \cdot \ve{x}}{x}\right)
\nonumber\\
&& +\frac{1}{4}\,(1 + \gamma)^2\,m^2\,
\frac{\ve{k}}{|\,\ve{x} \times \ve{x}_0\,|^2}\,
{R\over x}\,\left(1\,+\,\frac{\ve{k} \cdot \ve{x}}{x}\right)\,
(3x - x_0 - R)\,(x - x_0 + R)
\nonumber\\
&& + m^2\,\ve{k} \times ( \ve{x}_0 \times \ve{x} ) \,
\Bigg[ \,(1 + \gamma)^2\,\frac{R}{|\,\ve{x} \times \ve{x}_0\,|^2}
\,\left( 1 + \frac{\ve{k} \cdot \ve{x}}{x} \right)
\,\left(
\frac{R\,\left(R^2-(x-x_0)^2\right)}{2\,|\,\ve{x} \times \ve{x}_0\,|^2}
+{1\over x}\right)
\nonumber\\
&& -\frac{1}{2}\,\alpha\,\epsilon\,\frac{\ve{k} \cdot \ve{x}}{R\,x^4}
- {1\over 4}\,\left(8\,(1 + \gamma - \alpha\,\gamma)(1 + \gamma)
\,-4\,\alpha\,\beta\,+3\,\alpha\,\epsilon \right)
\frac{\ve{k} \cdot \ve{x}}{x^2}\,\frac{R}{|\,\ve{x} \times \ve{x}_0\,|^2}
\nonumber\\
&& - {1\over 4}\,
\left(8 (1 + \gamma-\alpha\,\gamma)(1 + \gamma)\,-4\,\alpha\,\beta\,
+ 3\,\alpha\,\epsilon \right)\,\frac{R^2}{|\,\ve{x} \times \ve{x}_0\,|^3}\,
\left( \pi - \delta (\ve{k} , \ve{x}) \right) 
\, \Bigg] + {\cal O}(c^{- 6})\,.
\nonumber\\
\label{n_10}
\end{eqnarray}

\noindent
This expression allows one to compute the difference between the vectors $\ve{n}$
and $\ve{\sigma}$, starting from the boundary conditions $\ve{x}_0$ and $\ve{x}$.

\subsection{Transformation of $\ve{k}$ to $\ve{n}$}

Finally, a direct relation between vectors $\ve{k}$ and $\ve{n}$ should be
derived. To this end, we directly combine
Eqs.~(\ref{k_to_sigma}) and  (\ref{n_10}) to get
\begin{eqnarray}
\ve{n} &=& {\ve{k}}
 - (1 + \gamma) \, m \,\frac{\ve{k} \times
(\ve{x}_0 \times \ve{x})}{x\left(x\,x_0 + \ve{x} \cdot \ve{x}_0\right)}\,(1 + F)
\nonumber\\
&& - \frac{1}{8}\,(1 + \gamma)^2\,\frac{m^2}{x^2}\,\ve{k}\,
\frac{{\left((x - x_0)^2 - R^2\right)}^2}{|\ve{x} \times \ve{x}_0|^2}
\nonumber\\
&& + \,m^2\, \ve{k} \times (\ve{x}_0 \times \ve{x})\,
\Biggl[
\,{1\over 2}\,(1 + \gamma)^2\,
\frac{R^2-(x-x_0)^2}{x^2\,|\ve{x} \times \ve{x}_0|^2}
\nonumber\\
&&  + \, \frac{1}{4} \, \alpha \, \epsilon \, \frac{1}{R}
\left(\frac{1}{R\,x_0^2} - \frac{1}{R\,x^2}
- 2\, \frac{\ve{k} \cdot \ve{x}}{x^4}\right)
\nonumber\\
&& - \frac{1}{4}\,\left(\, 8(1 + \gamma - \alpha \gamma) (1 + \gamma) - 4\alpha \beta
+ 3\, \alpha\, \epsilon \, \right)  \, R\,\frac{\ve{k} \cdot \ve{x}}
{x^2\,|\, \ve{x} \times \ve{x}_0 \,|^2}
\nonumber\\
&& + \frac{1}{8}\, \left(8 (1 + \gamma - \alpha \, \gamma) (1 + \gamma)
- 4 \, \alpha \, \beta + 3 \alpha \, \epsilon\right) \,
\frac{x^2 - x_0^2 - R^2}{|{\ve{x}} \times {\ve{x}}_0|^3} \,\delta(\ve{x} , \ve{x}_0)
\Biggr]
\, + \, {\cal O} \left( c^{- 6} \right)
\label{n_60}
\end{eqnarray}

\noindent
with
\begin{equation}
F=-(1 + \gamma) \,m\, \frac{x + x_0}{x\,x_0 + \ve{x} \cdot \ve{x}_0}\,.
\label{n_65}
\end{equation}

\noindent
This formula allows one directly to compute the unit coordinate direction
of light propagation $\ve{n}$ at the point of reception starting from
the positions of the source $\ve{x}_0$ and observer $\ve{x}$. The
reason for writing the post-post-Newtonian term proportional to $F$
together with the main post-Newtonian term is that it is the only
post-post-Newtonian term which cannot be estimated as ${\rm const}\cdot
m^2$. This will be discussed in a subsequent report.

\section{Conclusion}

The analytical post-post-Newtonian solution for light propagation
derived in this note will be used in a subsequent report to obtain an
analytical formula for light deflection for solar system bodies. Not
all post-post-Newtonian terms in the above formulas should be used to
obtain the goal accuracy of 1 \muas. Analytical estimations of
individual terms can be used to find those of them which are
numerically relevant at the level of 1 \muas. Detailed estimations and
comparison to numerical solutions of the boundary problem will be
given elsewhere.


\end{document}